\documentclass[aps,twocolumn]{revtex4}
\usepackage{graphicx}
\usepackage{epsfig}
\begin{document}
\bibliographystyle{apsrev}
\title{LISA Data Analysis using MCMC methods}
\author{Neil J. Cornish and Jeff Crowder}
\affiliation{Department of Physics, Montana State University, Bozeman, MT
59717}

\begin{abstract}
The Laser Interferometer Space Antenna (LISA) is expected to simultaneously detect many thousands of low
frequency gravitational wave signals. This presents a data analysis challenge that is very different to the
one encountered in ground based gravitational wave astronomy. LISA data analysis requires the identification
of individual signals from a data stream containing an unknown number of overlapping signals. Because of the
signal overlaps, a global fit to all the signals has to be performed in order to avoid biasing the solution.
However, performing such a global fit requires the exploration of an enormous parameter space with a
dimension upwards of 50,000. Markov Chain Monte Carlo (MCMC) methods offer a very promising solution to
the LISA data analysis problem. MCMC algorithms are able to efficiently explore large parameter spaces, simultaneously
providing parameter estimates, error analysis, and even model selection. Here we present the first application
of MCMC methods to simulated LISA data and demonstrate the great potential of the MCMC approach. Our implementation
uses a generalized F-statistic to evaluate the likelihoods, and simulated annealing to speed convergence of the
Markov chains. As a final step we super-cool the chains to extract maximum likelihood estimates, and estimates of
the Bayes factors for competing models. We find that the MCMC approach is able to correctly identify the number of
signals present, extract the source parameters, and return error estimates consistent with Fisher information
matrix predictions.
\end{abstract}

\maketitle

\section{Introduction}

The LISA observatory~\cite{lppa} has incredible science potential, but that potential can only be
fully realized by employing advanced data analysis techniques. LISA will explore the
low frequency portion of the gravitational wave spectrum, which is thought to
be home to a vast number of sources.
Since gravitational wave sources typically evolve on timescales
that are long compared to the gravitational wave period, individual low frequency sources will
be ``on'' for large fractions of the nominal three year LISA mission lifetime. Moreover,
unlike a traditional telescope, LISA can not be pointed at a particular point on the
sky. The upshot is that the LISA data stream will contain the signals from tens of thousands
of individual sources, and ways must be found to isolate individual voices from the crowd.
This ``Cocktail Party Problem'' is the central issue in LISA data analysis.

The types of sources LISA is expected to detect include galactic and extra-galactic
compact stellar binaries, super massive black hole binaries, and extreme mass ratio inspirals of
compact stars into supermassive black holes (EMRIs). Other potential sources include intermediate
mass black hole binaries, cosmic strings, and a cosmic gravitational wave background produced
by processes in the early universe. In the case of compact stellar
binaries~\cite{evans,lip,hils,hils2,gils} and EMRIs~\cite{cutbar,emri}, the number of sources is likely to be
so large that it will be impossible to resolve all the sources individually, so that there
will be a residual signal that is variously referred to as a confusion limited background
or confusion noise. It is important that this confusion noise be made as small as possible
so as not to hinder the detection of other high value targets. Several estimates of the
confusion noise level have been made~\cite{hils,hils2,gils,sterl,seth,bc}, and they all
suggest that unresolved signals will
be the dominant source of low frequency noise for LISA. However, these estimates are based
on assumptions about the efficacy of the data analysis algorithms that will be used to
identify and regress sources from the LISA data stream, and it is unclear at present how
reasonable these assumptions might be. Indeed, the very notion that one can first clean
the data stream of one type of signal before moving on to search for other targets is suspect
as the gravitational wave signals from different sources
are not orthogonal. For example, when the signal from a supermassive black hole
binary sweeps past the signal from a white dwarf binary of period $T$, the two signals will
have significant overlap for a time interval equal to the geometric mean of $T$ and $t_c$,
where $t_c$ is the time remaining before the black holes merge. Thus, by a process dubbed
``the white dwarf transform,'' it is possible to decompose the signal from a supermassive
black hole binary into signals from a collection of white dwarf binaries.

As described in \S\ref{cocktail}, optimal filtering of the LISA data would require the
construction of a filter bank  that described the signals from every source
that contributes to the data stream. In principle one could construct a vast template bank
describing all possible sources and look for the best match with the data. In practice
the enormous size of the search space and the presence of unmodeled sources renders this
direct approach impractical. Possible alternatives to a full template based search include
iterative refinement of a source-by-source search, ergodic exploration of the
parameter space using Markov Chain Monte Carlo (MCMC) algorithms , Darwinian optimization by genetic
algorithms, and global iterative refinement using the Maximum Entropy Method (MEM).
Each approach has its strengths and weakness, and at this stage it is not obvious which approach will
prove superior.

Here we apply the popular Markov Chain Monte Carlo~\cite{metro,haste} method to simulated LISA data. 
This is not the first time that MCMC methods have been applied to gravitational wave data analysis, but it is
first outing with realistic simulated LISA data. Our simulated data streams contain the signals from multiple
galactic binaries. Previously, MCMC methods have been used to study the extraction of coalescing binary~\cite{christ}
and spinning neutron star~\cite{woan} signals from terrestrial interferometers.
More recently, MCMC methods have been applied to a simplified toy
problem~\cite{woan2} that shares some of the features of the LISA cocktail party problem. These
studies have shown that MCMC methods hold considerable promise for gravitational wave data
analysis, and offer many advantages over the standard template grid searches. For example, the
EMRI data analysis problem~\cite{cutbar,emri} is often cited as the greatest challenge facing
LISA science. Neglecting the spin of the smaller body yields a 14 dimensional parameter space, which
would require $\sim 10^{40}$ templates to explore in a grid based search~\cite{emri}. This
huge computational cost arises because grid based searches scale geometrically with the parameter
space dimension $D$. In contrast, the computational cost of MCMC based searches scale linearly with
the $D$. In fields such as finance, MCMC methods are routinely applied to problems with
$D > 1000$, making the LISA EMRI problem seem trivial in comparison. A {\em Google} search on
``Markov Chain Monte Carlo'' returns almost 250,000 results, and a quick scan of these pages
demonstrates the wide range of fields where MCMC methods are routinely used. We found it amusing
that one of the {\em Google} search results is a
link to the {\em PageRank}~\cite{page} MCMC algorithm that powers the {\em Google} search engine.

The structure of the paper follows the development sequence we took to arrive at a fast and
robust MCMC algorithm. In \S\ref{cocktail} we outline the LISA data analysis problem and the
particular challenges posed by the galactic background. A basic MCMC algorithm is introduced in
\S\ref{MCMC7} and applied to a full 7 parameter search for a single galactic binary. A generalized
multi-channel, multi-source F-statistic for reducing the search space from $D=7 N$ to $D=3N$ is
described in \S\ref{Fstat}. The performance of a basic MCMC algorithm that uses the F-statistic is
studied in \S\ref{MCMC_F} and a number of problems with this simple approach are identified.
A more advanced mixed MCMC algorithm that incorporates simulated annealing is introduced in
\S\ref{MCMC_mix} and is successfully applied to multi-source searches. The issue of model selection
is addressed in \S\ref{bayes}, and approximate Bayes factor are calculated by super-cooling the
Markov Chains to extract maximum likelihood estimates. We conclude with a discussion of future
refinements and extensions of our approach in \S\ref{conclude}.

\section{The Cocktail Party Problem}\label{cocktail}

Space based detectors such as LISA are able to return several interferometer outputs~\cite{aet}.
The strains registered in the interferometer in response to a gravitational wave
pick up modulations due to the motion of the detector. The orbital motion
introduces amplitude, frequency, and phase modulation into the observed gravitational wave signal.
The amplitude modulation results from the detector's antenna pattern being
swept across the sky, the frequency modulation is due to the Doppler
shift from the relative motion of the detector and source, and the
phase modulation results from the detector's varying response to the
two gravitational wave polarizations~\cite{cc,cr}. These modulations encode information
about the location of the source. The modulations spread a monochromatic signal over
a bandwidth $\Delta f \sim (9+6(f/{\rm mHz})\sin\theta)f_m$, where $\theta$ is the co-latitude
of the source and $f_m=1/{\rm year}$ is the modulation frequency.
In the low frequency limit, where the wavelengths
are large compared to the armlengths of the detector, the interferometer
outputs $s_\alpha(t)$ can be combined to simulate the response of two independent 90 degree
interferometers, $s_I(t)$ and $s_{II}(t)$, rotated by 45 degrees with respect to each
other~\cite{cc,tom}. This allows LISA to measure both polarizations of the gravitational wave
simultaneously. A third combination of signals in the low frequency limit yields the symmetric
Sagnac variable~\cite{aet}, which is insensitive to gravitational waves and can be used to
monitor the instrument noise. When the wavelengths of the gravitational waves
become comparable to the size of the detector, which for LISA corresponds to frequencies above
10 mHz, the interferometry signals can be combined to give three independent time series with
comparable sensitivities~\cite{tom}.

The output of each LISA data stream can be written as
\begin{equation}\label{lisa_sig}
s_\alpha(t) = h_\alpha(t, \vec{\lambda}) + n_\alpha(t) 
            = \sum_{i=1}^N h^i_\alpha(t,\vec{\lambda}_i)+ n_\alpha(t) \, .
\end{equation}
Here $h^i_\alpha(t,\vec{\lambda}_i)$ describes the response registered in detector
channel $\alpha$ to a source with parameters $\vec{\lambda}_i$. The quantity
$h_\alpha(t,\vec{\lambda})$ denotes the combined response to a collection of $N$ sources
with total parameter vector $\vec{\lambda} = \sum_i \vec{\lambda}_i$ and
$n_\alpha(t)$ denotes the instrument noise in channel $\alpha$. Extracting the parameters
of each individual source from the combined response to all sources defines the
LISA cocktail party problem. In practice it will be impossible to resolve all of the
millions of signals that contribute to the LISA data streams. For one, there will not be enough
bits of information in the entire LISA
data archive to describe all $N$ sources in the Universe with signals that fall
within the LISA band. Moreover, most sources will produce signals that are well
below the instrument noise level, and even after optimal filtering most
of these sources will have signal to noise ratios below one. A more reasonable goal
might be to provide estimates for the parameters describing each of the $N'$ sources
that have integrated signal to noise ratios (SNR) above some threshold (such as
${\rm SNR} > 5$), where it is now understood that the noise includes the instrument
noise, residuals from the regression of bright sources, and the signals from unresolved sources.

While the noise will be neither
stationary nor Gaussian, it is not unreasonable to hope that the departures from
Gaussianity and stationarity will be mild. It is well know that matched filtering is the
optimal linear signal processing technique for signals with stationary Gaussian
noise~\cite{helstrom,wz}. Matched filtering is used extensively in all fields of
science, and is a popular data analysis technique in ground based gravitational wave
astronomy~\cite{kip,bernie,sathya1,sathya2,curt1,bala1,sathya3,ap1,eric1,sathya4,ben1,ben2}.
Switching to the Fourier domain, the signal
can be written as $\tilde{s}_\alpha(f) = \tilde{h}_{\alpha}(f,\vec{\lambda}')+\tilde{n}_{\alpha}(f)$,
where $\tilde{n}_{\alpha}(f)$ includes instrument noise and confusion noise, and the signals are described
by parameters $\vec{\lambda}'$. Using the standard noise weighted inner product for the independent data channels
over a finite observation time $T$,
\begin{equation}
( a \vert b ) = \frac{2}{T} \sum_\alpha \sum_f \frac{ \tilde{a}_\alpha^*(f) \tilde{b}_\alpha(f)
+ \tilde{a}_\alpha(f) \tilde{b}_\alpha^*(f)}
{S^{\alpha}_n(f)} \, ,
\end{equation}
a Wiener filter statistic can be defined:
\begin{equation}
\rho(\vec{\lambda}) = \frac{ ( s \vert h(\vec{\lambda}))}
{\sqrt{( h(\vec{\lambda})  \vert h(\vec{\lambda}))}} \, .
\end{equation}
The noise spectral density $S_n(f)$ is given in terms of the autocorrelation of the
noise
\begin{equation}
\langle n(f) n^*(f') \rangle = \frac{T}{2} \delta_{f f'} S_n(f) \, .
\end{equation}
Here and elsewhere angle brackets $\langle \rangle$ denote an expectation value.
An estimate for the source parameters $\vec{\lambda}'$ can be found by maximizing
$\rho(\vec{\lambda})$. If the noise is Gaussian and a signal is present, $\rho(\vec{\lambda})$ will be
Gaussian distributed with unit variance and mean equal to
the integrated signal to noise ratio
\begin{equation}
{\rm SNR} = \langle \rho(\vec{\lambda}') \rangle = \sqrt{( h(\vec{\lambda}')  \vert h(\vec{\lambda}'))}\, .
\end{equation}
The optimal filter for the LISA signal (\ref{lisa_sig})
is a matched template describing all $N'$ resolvable sources. The number of parameters
$d_i$ required to describe a source ranges from 7 for a slowly evolving circular galactic
binary to 17 for a massive black hole binary. A reasonable estimate~\cite{seth} for $N'$
is around $10^4$, so the full
parameter space has dimension $D = \sum_i d_i \sim 10^5$. Since the number of templates
required to uniformly cover a parameter space grows exponentially with $D$, a grid based
search using the full optimal filter is out of the question. Clearly an alternative
approach has to be found. Moreover, the number of resolvable sources $N'$ is not known
a priori, so some stopping criteria must be found to avoid over-fitting the data.

Existing approaches to the LISA cocktail party problem employ iterative schemes. The first
such approach was dubbed ``gCLEAN''~\cite{gclean} due to its similarity with the ``CLEAN''~\cite{Hoegbom}
algorithm that is used for astronomical image reconstruction. The ``gCLEAN'' procedure identifies and records
the brightest source that remains in the data stream, then subtracts a small amount of this source. The
procedure is iterated until a prescribed residual is reached, at which time the individual sources are
reconstructed from the subtraction record. A much faster
iterative approach dubbed ``Slice \& Dice''~\cite{slicedice} was recently proposed that proceeds by
identifying and fully subtracting the brightest source that remains in the data stream. A global
least squares re-fit to all the current list of sources is then performed, and the new
parameter record is used to produce a regressed data stream for the next iteration. Bayes factors
are used to provide a stopping criteria.

There is always the danger with iterative approaches that the procedure ``gets off on the wrong foot,''
and is unable to find its way back to the optimal solution. This can happen when two signals have
a high degree of overlap. A very different approach to the LISA source confusion problem is to
solve for all sources simultaneously using ergodic sampling techniques. Markov Chain Monte
Carlo (MCMC)~\cite{gilks,gamer} is a method for estimating the posterior
distribution, $p(\vec{\lambda} \vert s)$, that can be used with very large parameter spaces. The method is now
in widespread use in many fields, and is starting to be used by astronomers and cosmologists. One of the
advantages of MCMC is that it combines detection, parameter estimation, and the calculation of confidence intervals
in one procedure, as everything one can ask about a model is contained in $p(\vec{\lambda} \vert s)$.
Another nice feature of MCMC is that there are implementations that allow the number of parameters in
the model to be variable, with built in penalties for using too many parameters in the fit. In an MCMC
approach, parameter estimates from Wiener matched filtering are replaced by the Bayes estimator~\cite{davis}
\begin{equation}\label{be}
\lambda^i_{\rm B}(s) = \int \lambda^i \, p(\vec{\lambda} \vert s) \, d\vec{\lambda} \, ,
\end{equation}
which requires knowledge of $p(\vec{\lambda} \vert s)$ - the posterior distribution of
$\vec{\lambda}$ ({\it i.e.} the distribution of $\vec{\lambda}$ conditioned on the data $s$).
By Bayes theorem, the posterior distribution is related to the prior distribution $p(\vec{\lambda})$
and the likelihood $p(s \vert \vec{\lambda})$ by
\begin{equation}\label{post}
p(\vec{\lambda} \vert s) = \frac{ p(\vec{\lambda}) p(s \vert \vec{\lambda})}{ \int
p(\vec{\lambda'}) p(s \vert \vec{\lambda'}) d\vec{\lambda'} } \, .
\end{equation}
Until recently the Bayes estimator was little used in practical applications as the integrals
appearing in (\ref{be}) and (\ref{post}) are often analytically intractable. The traditional
solution has been to use approximations to the Bayes estimator, such as the maximum likelihood
estimator described below, however advances in the Markov Chain Monte Carlo technique allow direct
numerical estimates to be made.

When the noise $n(t)$ is a normal process with zero mean, the likelihood is given by~\cite{sam}
\begin{equation}\label{likely}
p(s \vert \vec{\lambda}) = C \exp\left[ -\frac{1}{2} \left( (s - h(\vec{\lambda})) \vert (s - h(\vec{\lambda}))
\right) \right]\, ,
\end{equation}
where the normalization constant $C$ is independent of $s$. In the large SNR limit the Bayes
estimator can be approximated by finding the dominant mode of the posterior distribution, $p(\vec{\lambda} \vert s)$,
which Finn~\cite{sam} and Cutler \& Flannagan\cite{curt1} refer to as a maximum likelihood estimator.
Other authors~\cite{kro,fstat} define the
maximum likelihood estimator to be the value of $\vec{\lambda}$ that maximizes the likelihood,
$p(s \vert \vec{\lambda})$. The former has the advantage of incorporating prior information, but the
disadvantage of not being invariant under parameter space coordinate transformations. The latter definition
corresponds to the standard definition used by most statisticians, and while it does not take into
account prior information, it is coordinate invariant. The two definitions give the same result for
uniform priors, and very similar results in most cases (the exception being where the priors have a large
gradient at maximum likelihood).

The standard definition of the likelihood yields an estimator that is identical to Wiener matched
filtering\cite{eche}. Absorbing normalization factors by adopting the inverted relative likelihood
${\cal L}(\vec{\lambda}) = p(s \vert 0)/p(s \vert \vec{\lambda})$, we have
\begin{equation}\label{ml}
\log {\cal L}(\vec{\lambda}) = (s\vert h(\vec{\lambda})) - \frac{1}{2} 
(h(\vec{\lambda})\vert h(\vec{\lambda})) \, .
\end{equation}
In the gravitational wave literature the quantity $\log {\cal L}(\vec{\lambda})$ is usually referred to
as the log likelihood, despite the inversion and rescaling. Note that
\begin{equation}\label{comp}
\langle \log {\cal L}(\vec{\lambda}') \rangle = \frac{1}{2} \langle \rho(\vec{\lambda}')\rangle^2
 = \frac{1}{2} \, {\rm SNR}^2 \, .
\end{equation}
The maximum likelihood estimator (MLE), $\vec{\lambda}_{\rm ML}$, is found by solving the coupled
set of equations $\partial \log {\cal L} / \partial \lambda^i = 0$. Parameter uncertainties can be
estimated from the negative Hessian of $\log {\cal L}$, which yields the Fisher Information Matrix
\begin{equation}
\Gamma_{ij}(\vec{\lambda}) = -{\Big \langle} \frac{\partial^2 \log {\cal L}(\vec{\lambda})}{ \partial \lambda^i
\partial \lambda^j} {\Big \rangle}  = (h_{, i} \vert h_{, j} ).
\end{equation}
In the large SNR limit the MLE can be found by writing $\vec{\lambda} = \vec{\lambda}' + \Delta\vec{\lambda}$
and Taylor expanding (\ref{ml}). Setting $\partial \log {\cal L} / \partial \Delta \lambda^i =0$ yields
the lowest order solution
\begin{equation}\label{mle}
\lambda^i_{\rm ML} = {\lambda'}^i + \Delta \lambda^i = {\lambda'}^i +\Gamma^{ij}(\vec{\lambda'})(n \vert h_{, j}) \, .
\end{equation}
The expectation value of the maximum of the log likelihood is then
\begin{equation}\label{max}
\langle \log {\cal L}(\vec{\lambda}_{\rm ML})\rangle = \frac{{\rm SNR}^2 + D}{2} \, .
\end{equation}
This value exceeds that found in (\ref{comp}) by an amount that depends on the total number of parameters
used in the fit, $D$, reflecting the fact that models with more
parameters generally give better fits to the data. Deciding how many parameters to allow in the fit
is an important issue in LISA data analysis as the number of resolvable sources is not
known a priori. This issue does not usually arise for ground based gravitational wave
detectors as most high frequency gravitational wave sources are transient. The relevant
question there is whether or not a gravitational wave signal is present in a section of the data
stream, and this question can be dealt with by the Neyman-Pearson test or other similar
tests that use thresholds on the likelihood ${\cal L}$ that are related to the false
alarm and false dismissal rates. Demanding that ${\cal L} > 1$ - so it is more likely that
a signal is present than not - and setting a detection threshold of $\rho = 5$ yields
a false alarm probability of 0.006 and a detection probability of 0.994 (if the noise
is stationary and Gaussian). A simple acceptance threshold of $\rho = 5$ for each individual
signal used to fit the LISA data would help restrict the total number of parameters in the fit,
however there are better criteria that can be employed.
The simplest is related to the Neyman-Pearson test and compares the likelihoods
of models with different numbers of parameters. For nested models this ratio has an approximately
chi squared distribution which allows the significance of adding extra parameters to be
determined from standard statistical tables. A better approach is to compute the Bayes factor, 
\begin{equation}
B_{XY} = \frac{p_X(s)}{p_Y(s)} \, ,
\end{equation}
which gives the relative weight of evidence for models $X$ and $Y$ in terms of the ratio of marginal
likelihoods
\begin{equation}\label{marginal}
p_X(s) = \int p(s \vert \vec{\lambda},X) p(\vec{\lambda},X) d\vec{\lambda} \, .
\end{equation}
Here $p(s \vert \vec{\lambda},X)$ is the likelihood distribution for model $X$ and
$p(\vec{\lambda},X)$ is the prior distribution for model $X$. The difficulty with
this approach is that the integral in (\ref{marginal}) is hard to calculate, though
estimates can be made using the Laplace approximation or the Bayesian Information Criterion (BIC)~\cite{schwarz}.
The Laplace approximation is based on the method of steepest descents, and for uniform
priors yields
\begin{equation}
p_X(s) \simeq p(s \vert \vec{\lambda}_{\rm ML},X) \left( \frac{\Delta V_X}{V_X} \right) \, ,
\end{equation}
where $p(s \vert \vec{\lambda}_{\rm ML},X)$ is the maximum likelihood for the model, $V_X$ is the volume
of the model's parameter space, and $\Delta V_X$ is the volume of the
uncertainty ellipsoid (estimated using the Fisher matrix). Models with more parameters generally
provide a better fit to the data and a higher
maximum likelihood, but they get penalized by the $\Delta V_X/V_X$ term which acts as a built in Occam's razor.

\section{Markov Chain Monte Carlo}\label{MCMC7}

We begin by implementing a basic MCMC search for galactic binaries that searches over the
full $D=7N$ dimensional parameter space using the Metropolis-Hastings~\cite{haste} algorithm.
The idea is to generate a set of samples, $\{ \vec{x} \}$, that correspond to draws from the
posterior distribution, $p(\vec{\lambda} \vert s)$. To do this we start at a randomly chosen
point $\vec{x}$ and generate a Markov chain according to the following algorithm:
Using a proposal distribution $q(\cdot \vert \vec{x})$, draw a new point
$\vec{y}$. Evaluate the Hastings ratio
\begin{equation}
H = \frac{p(\vec{y}) p(s \vert \vec{y}) q(\vec{x} \vert \vec{y})}
{p(\vec{x}) p(s \vert \vec{x}) q(\vec{y} \vert \vec{x})} \, .
\end{equation}
Accept the candidate point $\vec{y}$ with probability $\alpha = {\rm min}(1,H)$, otherwise remain
at the current state $\vec{x}$ (Metropolis rejection~\cite{metro}). Remarkably, this
sampling scheme produces a Markov chain with a stationary distribution equal to the posterior
distribution of interest, $p(\vec{\lambda} \vert s)$, regardless of the choice of proposal
distribution~\cite{gilks}. A concise introduction to MCMC methods can be found in the review
paper by Andrieu {\it et al}~\cite{mcmc_hist}.

On the other hand, a poor choice of the proposal distribution will result in the algorithm taking
a very long time to converge to the stationary distribution (known as the burn-in time). 
Elements of the Markov chain produced during the burn-in phase have to be discarded as they do not
represent the stationary distribution. When dealing
with large parameter spaces the burn-in time can be very long if poor techniques are used. For
example, the Metropolis sampler, which uses symmetric proposal distributions, explores the parameter
space with an efficiency of at most $\sim 0.3/D$, making it a poor choice for high
dimension searches. Regardless of the sampling scheme, the mixing of the Markov chain can be
inhibited by the presence of strongly correlated parameters. Correlated parameters can be dealt with by
making a local coordinate transformation at $\vec{x}$ to a new set of coordinates that
diagonalises the Fisher matrix, $\Gamma_{ij}(\vec{x})$.

We tried a number of proposal distributions and update schemes to search for a single galactic binary.
The results were very disappointing. Bold proposals that attempted large jumps had a very poor acceptance
rate, while timid proposals that attempted small jumps had a good acceptance rate, but they explored the
parameter space very slowly, and got stuck at local modes of the posterior. Lorentzian proposal distributions fared
the best as their heavy tails and concentrated peaks lead to a mixture of bold and timid jumps, but
the burn in times were still very long and the subsequent mixing of the chain was torpid. The MCMC literature
is full of similar examples of slow exploration of large parameter spaces, and a host of schemes have
been suggested to speed up the burn-in. Many of the accelerated algorithms use adaptation to tune the
proposal distribution. This violates the Markov nature of the chain as the updates depend
on the history of the chain. More complicated adaptive algorithms have been invented that restore the Markov
property by using additional Metropolis rejection steps. The popular Delayed Rejection Method~\cite{dr} and Reversible
Jump Method~\cite{rj} are examples of adaptive MCMC algorithms. A simpler approach is to use a non-Markov scheme
during burn-in, such as adaptation or simulated annealing, then transition to a Markov scheme after
burn-in. Since the burn-in portion of the chain is discarded, it does not matter if the MCMC rules are
broken (the burn-in phase is more like Las Vegas than Monte Carlo). 

Before resorting to complex acceleration schemes we tried a much simpler approach that proved to
be very successful. When using the Metropolis-Hastings algorithm there is no reason to restrict
the updates to a single proposal distribution. For example, every update could use a different
proposal distribution so long as the choice of distribution is not based on the history of
the chain. The proposal distributions to be used at each update can be chosen at random, or they
can be applied in a fixed sequence. Our experience with single proposal distributions suggested
that a scheme that combined a very bold proposal with a very timid proposal would lead to fast
burn-in and efficient mixing. For the bold proposal we chose a uniform distribution for each of
the source parameters $\vec{\lambda} \rightarrow (A,f,\theta, \phi, \psi, \iota, \varphi_0)$.
Here $A$ is the amplitude, $f$ is the gravitational wave frequency, $\theta$ and $\phi$ are the
ecliptic co-latitude and longitude, $\psi$ is the polarization angle, $\iota$ is the inclination
of the orbital plane, and $\varphi_0$ is the orbital phase at some fiducial time.
The amplitudes were restricted to the range $A \in [10^{-23}, 10^{-21}]$ and the frequencies
were restricted to lie within the range of the data snippet $f \in [0.999995,1.003164]$ mHz
(the data snippet contained 100 frequency bins of width $\Delta f = 1/{\rm year}$). A better
choice would have been to use a cosine distribution for the co-latitude $\theta$ and inclination
$\iota$, but the choice is not particularly important. When multiple sources were present each source
was updated separately during the bold proposal stage.
For the timid proposal we used a normal distribution for
each eigendirection of the Fisher matrix, $\Gamma_{ij}(\vec{x})$. The standard deviation $\sigma_{\hat{k}}$
for each eigendirection $k$ was set equal to $\sigma_{\hat{k}} = 1/\sqrt{\alpha_{\hat{k}} D}$, where
$\alpha_{\hat{k}}$ is the corresponding eigenvalue of $\Gamma_{ij}(\vec{x})$, and $D=7N$ is the
search dimension. The factor of $1/\sqrt{D}$ ensures a healthy acceptance rate as the typical total jump
is then $\sim 1\sigma$. All $N$ sources were updated simultaneously during the timid proposal stage.
Note that the timid proposal distributions are not symmetric since $\Gamma_{ij}(\vec{x}) \neq
\Gamma_{ij}(\vec{y})$. One set of bold proposals (one for each source) was followed by ten timid
proposals in a repeating cycle. The ratio of the number of bold to timid proposals impacted the burn-in times
and the final mixing rate, but ratios anywhere from 1:1 to 1:100 worked well. We used uniform priors,
$p(\vec{x}) = {\rm const.}$, for all the parameters, though once again a cosine distribution would
have been better for $\theta$ and $\iota$. Two independent LISA data channels were simulated directly
in the frequency domain using the method described in Ref.~\cite{seth}, with the sources chosen at
random using the same uniform distributions employed by the bold proposal. The
data covers 1 year of observations, and the data snippet contains 100 frequency bins (of width
$1/{\rm year}$). The
instrument noise was assumed to be stationary and Gaussian, with position noise spectral density
$S_n^{\rm pos} = 4 \times 10^{-22}\; {\rm m}^2 {\rm Hz}^{-1}$ and acceleration noise spectral
density $S_n^{\rm accel} = 9 \times 10^{-30}\; {\rm m}^2 {\rm s}^{-4} {\rm Hz}^{-1}$.

\begin{table}[t]
\caption{7 parameter MCMC search for a single galactic binary}
\begin{tabular}{|l|c|c|c|c|c|c|c|}
\hline
   & $A$ ($10^{-22}$) & $f$ (mHz) & $\theta$ & $\phi$ & $\psi$ & $\iota$ & $\varphi_0$ \\
 \hline 
$\vec{\lambda}_{\rm True}$  & 1.73 & 1.0005853 & 0.98 & 4.46 & 2.55 & 1.47 & 0.12 \\
$\vec{\lambda}_{\rm MCMC}$  & 1.44 &  1.0005837 & 1.07 & 4.42 & 2.56 & 1.52 & 0.15 \\
\hline 
$\sigma_{\rm Fisher}$  & 0.14 & 2.2e-06 & 0.085 & 0.051 & 0.054 & 0.050 & 0.22 \\
$\sigma_{\rm MCMC}$  & 0.14 & 2.4e-06 & 0.089 & 0.055 & 0.058 & 0.052 & 0.23 \\
\hline  
\end{tabular}
\label{tab1}
\end{table}

\begin{figure}[h]
\includegraphics[angle=0,width=0.5\textwidth]{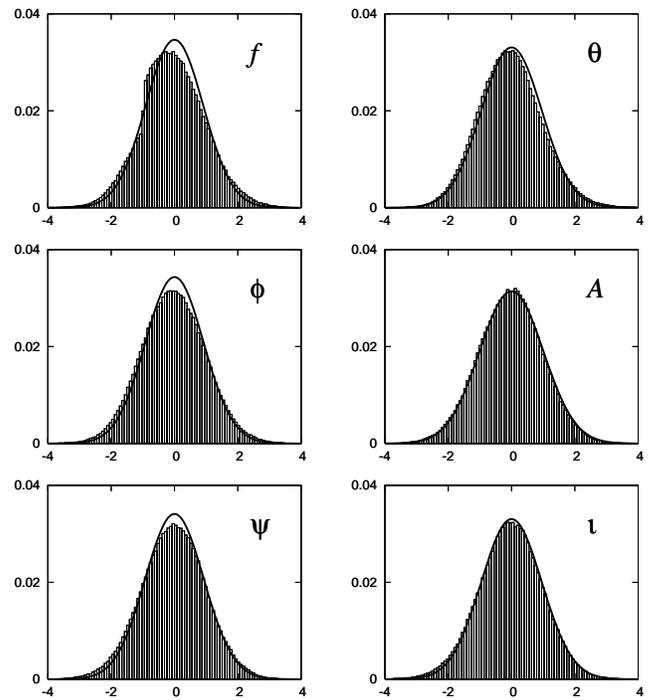}
\caption{\label{MCMC7_fig}Histograms showing the posterior distribution (grey) of the parameters. Also shown (black line)
is the Gaussian approximation to the posterior distribution based on the Fisher information matrix. The mean
values have been subtracted, and the parameters have been scaled by the square root of the variances calculated from
the MCMC chains.}
\end{figure}

Table~\ref{tab1} summarizes the results of one MCMC run using a model with one source to search for a
single source in the data snippet. Burn-in lasted $\sim 2000$ iterations, and post burn-in the chain was
run for $10^6$ iterations with a proposal acceptance rate of $77\%$ (the full run took 20 minutes on
a Mac G5 2 GHz processor). The chain was used to calculate means and variances for all the parameters.
The parameter uncertainty estimates extracted from the MCMC output are compared to the Fisher matrix
estimates evaluated at the mean values of the parameters. The source had true ${\rm SNR}=12.9$, and MCMC
recovered ${\rm SNR}=10.7$. Histograms of the posterior parameter distributions are
shown in Figure~\ref{MCMC7_fig}, where they are compared to the Gaussian approximation to the posterior given by the
Fisher matrix. The agreement is impressive, especially considering that the bandwidth of the source
is roughly 10 frequency bins, so there are very few noise samples to work with. Similar results were
found for other MCMC runs on the same source, and for MCMC runs with other sources. Typical burn-in times
were of order 3000 iterations, and the proposal acceptance rate was around $75\%$.

The algorithm was run successfully on two and three source searches (the model dimension was chosen to
match the number of sources in each instance), but on occasions the chain would get stuck at a local
mode of the posterior for a large number of iterations. Before attempting to cure this problem with
a more refined MCMC algorithm, we decided to eliminate the extrinsic parameters $A, \iota, \psi, \varepsilon_0$
from the search by using a multi-filter generalized F-statistic. This reduces the search dimension to
$D=3N$, with the added benefit that the projection onto the $(f, \theta, \phi)$ sub-space yields a softer
target for the MCMC search.

\section{Generalized F-Statistic}\label{Fstat}

The F-statistic was originally introduced~\cite{fstat} in the context of ground based searches for
gravitational wave signals from rotating Neutron stars. The F-statistic has since been used
to search for monochromatic galactic binaries using simulated LISA data~\cite{flisa, slicedice}.
By using multiple linear filters, the F-statistic is able to automatically extremize the
log likelihood over extrinsic parameters, thus reducing the dimension of the search space (the
parameter space dimension remains the same).

In the low-frequency limit the LISA response to a gravitational wave with polarization content $h_+(t)$,
$h_\times(t)$ can be written as
\begin{equation}\label{resp}
h(t) = h_+(t) F^+(t) +  h_\times(t) F^\times(t) \, ,
\end{equation}
where
\begin{eqnarray}
F^+(t) &=& \frac{1}{2} \left( \cos 2 \psi \, D^+(t) - \sin 2 \psi \, D^\times(t)\right) \nonumber \\
F^\times(t) &=& \frac{1}{2} \left( \sin 2 \psi \, D^+(t) + \cos 2 \psi \, D^\times(t)\right)
\end{eqnarray}
The detector pattern functions $D^+(t)$ and $D^\times(t)$ are given in equations (36) and (37) of
Ref.\cite{rigad}. To leading post-Newtonian order a slowly evolving, circular binary has polarization
components
\begin{eqnarray}
h_+(t) &=& A (1+\cos^2\iota) \cos(\Phi(t)+\varphi_0) \nonumber \\
h_\times(t) &=& -2 A \cos\iota\, \sin(\Phi(t)+\varphi_0).
\end{eqnarray}
The gravitational wave phase
\begin{equation}
\Phi(t; f, \theta, \phi) = 2\pi f t 
+2 \pi f {\rm AU} \sin\theta \cos(2 \pi f_m t - \phi),
\end{equation}
couples the sky location and the frequency through the term that depends on the radius of LISA's orbit, 1 AU, and
the orbital modulation frequency, $f_m = 1/{\rm year}$. The gravitational wave amplitude, $A$, is
effectively constant for the low frequency galactic sources we are considering.
Using these expressions (\ref{resp}) can be written as
\begin{equation}\label{aA}
h(t) = \sum_{i=1}^4 a_i(A,\psi,\iota,\varphi_0) A^i(t; f, \theta, \phi)\, ,
\end{equation}
where the time-independent amplitudes $a_i$ are given by
\begin{eqnarray}
a_1 &=& \frac{A}{2} \left( (1+\cos^2\iota)\cos\varphi_0 \cos 2\psi-2\cos\iota \sin\varphi_0 \sin 2 \psi \right),
\nonumber \\
a_2 &=& -\frac{A}{2} \left( 2\cos\iota\sin\varphi_0 \cos 2\psi+ (1+\cos^2\iota) \cos\varphi_0 \sin 2 \psi \right),
\nonumber \\
a_3 &=& -\frac{A}{2} \left( 2\cos\iota\cos\varphi_0 \sin 2\psi+ (1+\cos^2\iota) \sin\varphi_0 \cos 2 \psi \right),
\nonumber \\
a_4 &=& \frac{A}{2} \left( (1+\cos^2\iota)\sin\varphi_0 \sin 2\psi-2\cos\iota \cos\varphi_0 \cos 2 \psi \right),
\nonumber \\
\end{eqnarray}
and the time-dependent functions $A^i(t)$ are given by
\begin{eqnarray}
A^1(t) &=& D^+(t;\theta, \phi)  \cos \Phi(t;f, \theta, \phi)
\nonumber \\
A^2(t) &=& D^\times(t;\theta, \phi) 
\cos \Phi(t;f, \theta, \phi) \nonumber \\
A^3(t) &=& D^+(t;\theta, \phi) 
\sin \Phi(t;f, \theta, \phi) \nonumber \\
A^4(t) &=& D^\times(t;\theta, \phi) 
\sin \Phi(t;f,\theta, \phi) \, .
\end{eqnarray}
Defining the four constants $N^i=(s \vert A^i)$
and using (\ref{aA}) yields a solution for the amplitudes $a_i$:
\begin{equation}\label{asol}
a_i = (M^{-1})_{ij} N^j \, ,
\end{equation}
where $M^{ij}=(A^i\vert A^j)$. The output of the four linear filters, $N^i$, and the
$4\times 4$ matrix $M^{ij}$ can be calculated using the same fast Fourier space techniques~\cite{seth}
used to generate the full waveforms. Substituting (\ref{aA}) and (\ref{asol}) into
expression (\ref{ml}) for the log likelihood yields the F-statistic
\begin{equation}\label{fstat}
{\cal F} = \log {\cal L}=\frac{1}{2} (M^{-1})_{ij} N^i N^j \, .
\end{equation}
The F-statistic automatically maximizes the log likelihood over the
extrinsic parameters $A,\, \iota, \, \psi$ and $\varphi_0$, and reduces the search to
the sub-space spanned by $f,\, \theta$ and $\phi$. The extrinsic parameters can be
recovered from the $a_i$'s via
\begin{eqnarray}
A & = & \frac{ A_+ + \sqrt{A_+^2-A_\times^2}}{2} \nonumber \\
\psi & =& \frac{1}{2}\arctan\left(\frac{A_+ a_4 - A_\times a_1}{-(A_\times a_2 + A_+ a_3)}\right) \nonumber \\
\iota & = & \arccos\left(\frac{-A_\times}{A_+ + \sqrt{A_+^2-A_\times^2}}\right)\nonumber \\
\varphi_0 & =& \arctan\left(\frac{c(A_+ a_4 - A_\times a_1)}{-c(A_\times a_2 + A_+ a_3)}\right)
\end{eqnarray}
where
\begin{eqnarray}
A_+ &=& \sqrt{(a_1+a_4)^2 + (a_2-a_3)^2} \nonumber \\
&& + \sqrt{(a_1-a_4)^2 + (a_2+a_3)^2} \nonumber \\
A_\times &=& \sqrt{(a_1+a_4)^2 + (a_2-a_3)^2} \nonumber \\
&& - \sqrt{(a_1-a_4)^2 + (a_2+a_3)^2} \nonumber \\
c &=& {\rm sign}(\sin(2 \psi)) \, .
\end{eqnarray}

The preceding description of the
F-statistic automatically incorporates the two independent LISA channels through
the use of the dual-channel noise weighted inner product $(a\vert b)$. The basic
F-statistic can easily be generalized to handle $N$ sources. Writing
$i= 4K +l$, where $K$ labels the source and $l = 1 \rightarrow 4$ labels the four filters
for each source, the F-statistic (\ref{fstat}) keeps the same form as before, but now
there are $4N$ linear filters $N^i$, and $M^{ij}$ is a $4N \times 4N$ dimensional matrix. For slowly
evolving galactic binaries, which dominate the confusion problem, the limited bandwidth of each
individual signal means that the $M^{ij}$ is band diagonal, and thus easily inverted despite its
large size.

Since the search is now over the projected sub-space $\{f_J, \theta_J, \phi_J\}$ of the full
parameter space, the full Fisher matrix, $\Gamma_{ij}(\vec{x})$, is replaced by the projected
Fisher matrix, $\gamma_{ij}(\vec{x})$. The projection of the $k^{\rm th}$ parameter is given by
\begin{equation}
\Gamma^{n-1}_{ij} = \Gamma^{n}_{ij} - \frac{\Gamma^n_{ik}\Gamma^n_{jk}}{\Gamma^n_{kk}} \, ,
\end{equation}
where $n$ denotes the dimension of the projected matrix. Repeated application of the above
projection yields $\gamma_{ij}=\Gamma^{3N}_{ij}$. Inverting $\gamma_{ij}$ yields the same
uncertainty estimates for the intrinsic parameters as one gets from the full Fisher matrix, but the
covariances are much larger. The large covariances make it imperative that the proposal
distributions use the eigenvalues and eigenvectors of $\gamma_{ij}$, as using the parameter
directions themselves would lead to a slowly mixing chain.

\section{F-Statistic MCMC}\label{MCMC_F}

\begin{figure}[h]
\includegraphics[angle=0,width=0.5\textwidth]{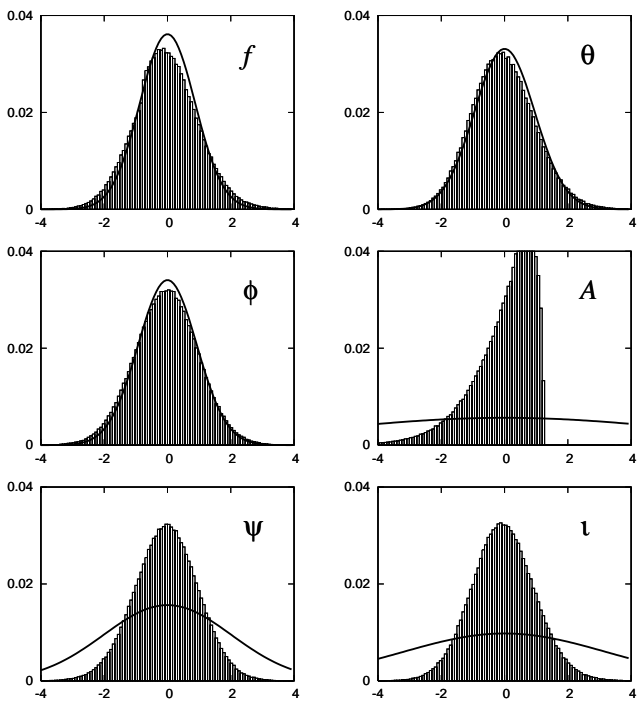}
\caption{\label{FMCMC_fig}Histograms showing the posterior distribution (grey) of the parameters. Also shown (black line)
is the Gaussian approximation to the posterior distribution based on the Fisher information matrix. The mean
values have been subtracted, and the parameters have been scaled by the square root of the variances calculated from
the F-MCMC chains.}
\end{figure}

We implemented an F-statistic based MCMC algorithm using the approach described in
\S\ref{MCMC7}, but with the full likelihood replaced by the F-statistic and the full Fisher
matrix replaced by the projected Fisher matrix. Applying the F-MCMC search to the same
data set as before yields the results summarized in Figure~\ref{FMCMC_fig} and Table~\ref{tab2}.
The recovered source parameters and signal-to-noise ratio (${\rm SNR} = 10.4$) are very similar
to those found using the full 7-parameter
search, but the F-MCMC estimates for the errors in the extrinsic parameters are very
different. This is because the chain does not explore extrinsic parameters, but rather
relies upon the F-statistic to find the extrinsic parameters that give the largest
log likelihood based on the current values for the intrinsic parameters. The effect is very
pronounced in the histograms shown in Figure~\ref{FMCMC_fig}. Similar results were
found for other F-MCMC runs on the same source, and for F-MCMC runs with other sources. Typical
burn-in times were of order 1000 iterations, and the proposal acceptance rate was around $60\%$.
As expected, the F-MCMC algorithm gave shorter burn-in times than the full parameter MCMC, and
a comparable mixing rate.

\begin{table}[t]
\caption{F-MCMC search for a single galactic binary}
\begin{tabular}{|l|c|c|c|c|c|c|c|}
\hline
   & $A$ ($10^{-22}$) & $f$ (mHz) & $\theta$ & $\phi$ & $\psi$ & $\iota$ & $\varphi_0$ \\
 \hline 
$\vec{\lambda}_{\rm True}$  & 1.73 & 1.0005853 & 0.98 & 4.46 & 2.55 & 1.47 & 0.12 \\
$\vec{\lambda}_{\rm F-MCMC}$  & 1.38 & 1.0005835 & 1.09 & 4.42 & 2.56 & 1.51 & 0.17 \\
\hline 
$\sigma_{\rm Fisher}$  &  0.14 & 2.2e-06 & 0.089 & 0.052 & 0.055 & 0.051 & 0.22 \\
$\sigma_{\rm MCMC}$  &  0.02 & 2.5e-06 & 0.093 & 0.056 & 0.027 & 0.016 & 0.21  \\
\hline  
\end{tabular}
\label{tab2}
\end{table}

It is interesting to compare the computational cost of the F-MCMC search to a traditional
F-Statistic based search on a uniformly spaced template grid. To cover the parameter
space of one source (which for the current example extends over the full sky and 100 frequency bins)
with a minimal match~\cite{ben1} of ${\rm MM}=0.9$ requires 39,000 templates~\cite{neil_ed}.
A typical F-MCMC run uses less than 1000 templates to cover the same search space. The
comparison becomes even more lopsided if we consider simultaneous searches for multiple
sources. A grid based simultaneous search for two sources using the F-statistic would take
$(39,000)^2\simeq 1.5 \times 10^9$ templates, while the basic F-MCMC algorithm typically
converges on the two sources in just 2000 steps. As the number of sources in the model
increases the computation cost of the grid based search grows geometrically while the
cost of the F-MCMC search grows linearly. It is hard to imagine a scenario (other than
quantum computers) where non-iterative grid based searches could play a role in LISA
data analysis.

\begin{figure}[h]
\includegraphics[angle=0,width=0.45\textwidth]{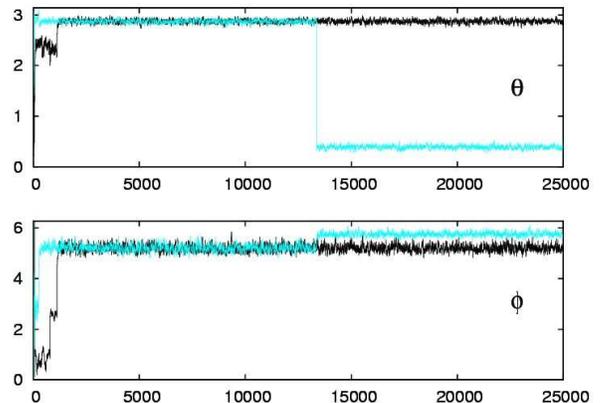}
\caption{\label{FMCMC_trace}Trace plots of the sky location parameters for two F-MCMC runs on the same data set.
Both chains initially locked onto a secondary mode of the posterior, but one of the chains (light colored line)
transitioned to the correct mode after 13,000 iterations.}
\end{figure}

\begin{figure}[h]
\includegraphics[angle=0,width=0.5\textwidth]{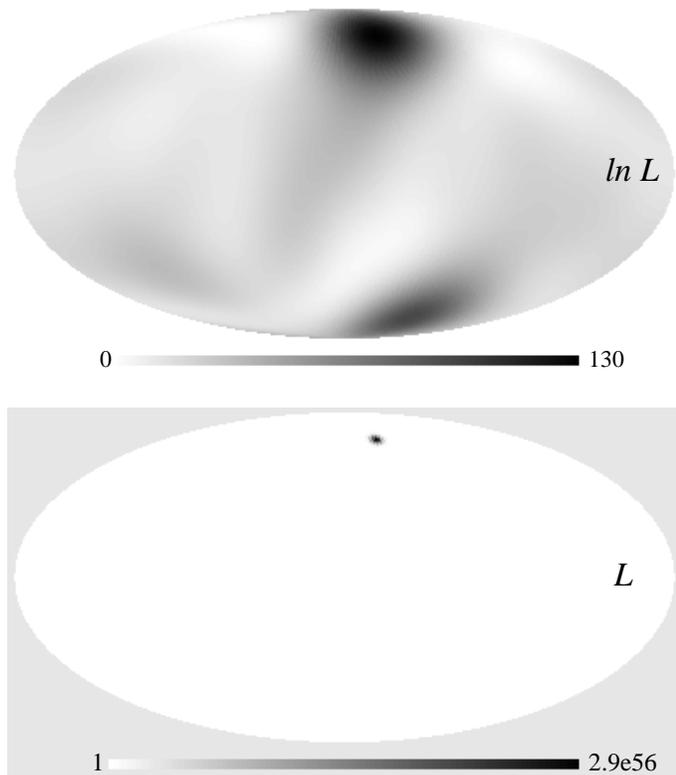}
\caption{\label{sky}Inverted likelihood and log likelihood as a function of sky location at fixed frequency.}
\end{figure}

While testing the F-MCMC algorithm on different sources we came across instances where the chain became
stuck at secondary modes of the posterior. A good example occurred for a source with parameters
$(A,f,\theta,\phi,\psi,\iota,\varphi_0)$=$(1.4$e-22$,1.0020802\, {\rm mHz},$ $0.399, 5.71, 1.3, 0.96, 1.0)$
and ${\rm SNR} = 16.09$. Most MCMC runs returned good fits to the source parameters, with an average
log likelihood of $\ln{\cal L} = 132$, mean intrinsic parameter values
$(f,\theta,\phi)= (1.0020809\, {\rm mHz}, 0.391, 5.75)$ and ${\rm SNR} = 16.26$. However, some
runs locked into a secondary mode with average log likelihood $\ln{\cal L} = 100$, mean intrinsic
parameter values $(f,\theta,\phi)= (1.0020858\, {\rm mHz}, 2.876, 5.20)$ and ${\rm SNR} = 14.15$.
It could sometimes take hundreds of thousands of iterations for the chain to discover the dominant
mode. Figure~\ref{sky} shows plots of the (inverted) likelihood ${\cal L}$ and the log likelihood $\ln {\cal L}$
as a function of sky location for fixed $f=1.0020802\, {\rm mHz}$. The log likelihood plot
reveals the problematic secondary mode near the south pole, while the likelihood plot shows just how
small a target the dominant mode presents to the F-MCMC search. Similar problems with secondary modes
were encountered in the $f-\phi$ plane, where the chain would get stuck a full bin away
from the correct frequency. These problems with the basic F-MCMC algorithm motivated the embellishments
described in the following section.

\section{Multiple Proposals and Heating}\label{MCMC_mix}

The LISA data analysis problem belongs to a particularly challenging class of MCMC problems
known as ``mixture models.'' As the name suggests, a mixture model contains a number of
components, some or all of which may be of the same type. In our present study all the components
are slowly evolving, circular binaries, and each component is described by the same set of
seven parameters. There is nothing to stop two components in the search model from latching on
to the same source, nor is there anything to stop one component in the search model from
latching on to a blend of two overlapping sources. In the former instance the likelihood is little
improved by using two components to model one source, so over time one of the components will tend to
wander off in search of another source. In the latter instance it may prove impossible for
any data analysis method to de-blend the sources (the marginal likelihood for the single component fit
to the blended sources may exceed the marginal likelihood of the ``correct'' solution).

The difficulties we encountered with the single source searches getting stuck at secondary modes of the
posterior are exacerbated in the multi-source case. Source overlaps can create additional secondary modes
that are not present in the non-overlapping case. We employed two techniques to speed burn-in and to
reduce the chance of the chain getting stuck at a secondary mode: simulated annealing and multiple
proposal distributions. Simulated annealing works by softening the likelihood function, making it
easier for the chain to move between modes. The likelihood (\ref{likely}) can be thought of as
a partition function $Z = C \exp(-\beta E)$ with the ``energy'' of the system given by
$E = (s-h \vert s-h)$ and the ``inverse temperature'' equal to $\beta = 1/2$. Our goal is to find
the template $h$ that minimizes the energy of the system. Heating up the system by setting $\beta < 1/2$
allows the Markov Chain to rapidly explore the likelihood surface. We used a standard power law cooling
schedule:
\begin{equation}
\beta = \left\{
\begin{tabular}{ll}
$\beta_0 \left(\frac{1}{2\beta_0}\right)^{t/T_c}$ & $0 < t < T_c$ \\
$\frac{1}{2}$ &  $t \geq T_c $
\end{tabular}
\right.
\end{equation}
where $t$ is the number of steps in the chain, $T_c$ is the cooling time and $\beta_0$ is
the initial inverse temperature. It took some trial and error to find good values of
$T_c$ and $\beta_0$. If some of the sources have very high SNR it is a good idea to start
at a high temperature $\beta_0 \sim 1/50$, but in most cases we found $\beta_0 = 1/10$ to
be sufficient. The optimal choice for the cooling time depends on the number of sources
and the initial temperature. We found that it was necessary to increase $T_c$ roughly linearly with
the the number of sources and the initial temperature. Setting $T_c = 10^5$ for a model with
$N=10$ sources and an initial temperature of $\beta_0 = 1/10$ gave fairly reliable results, but
it is always a good idea to allow longer cooling times if the computational resources are
available. The portion of the chain generated during the annealing phase has to be discarded
as the cooling introduces an arrow of time which necessarily violates the reversibility
requirement of a Markov Chain.

After cooling to $\beta=1/2$ the chain can explore the likelihood surface
for the purpose of extracting parameter estimates and error estimates. Finally, we can extract
maximum likelihood estimates by ``super cooling'' the chain to some very low temperature
(we used $\beta \sim 10^4$).

The second ingredient in our advanced F-MCMC algorithm is a large variety of
proposal distributions. We used the following types of proposal distribution:
${\rm Uniform}(\cdot, \vec{x}, i)$ - a uniform draw on all the parameters that
describe source $i$, using the full parameter ranges, with all other sources held fixed;
${\rm Normal}(\cdot, \vec{x})$ - a multivariate normal distribution with variance-covariance matrix given by
$3N\times \gamma(\vec{x})$; ${\rm Sky}(\cdot, \vec{x}, i)$ - a uniform draw on the sky
location for source $i$; $\sigma$-Uniform$(\cdot, \vec{x}, i)$ - a uniform draw on all the parameters that
describe source $i$, using a parameter range given by some multiple of the standard deviations
given by $\gamma(\vec{x})$. The ${\rm Uniform}(\cdot, \vec{x}, i)$ and ${\rm Normal}(\cdot, \vec{x})$
proposal distributions are the same as those used in the basic F-MCMC algorithm. The ${\rm Sky}(\cdot, \vec{x}, i)$
proposal proved to be very useful at getting the chain away from secondary modes like the one seen in
Figure~\ref{sky}, while the $\sigma$-Uniform$(\cdot, \vec{x}, i)$ proposal helped to
move the chain from secondary modes in the $f-\phi$ or $f-\theta$ planes. During the initial
annealing phase the various proposal distributions were used in a cycle with one set of
the bold distributions (Uniform, Sky and $\sigma-$Uniform) for every 10 draws from the timid multivariate
normal distribution. During the main MCMC run at $\beta=1/2$ the ratio of timid to bold proposals
was increased by a factor of 10, and in the final super-cooling phase only the timid multivariate normal
distribution was used.

The current algorithm is intended to give a proof of principle, and is
certainly far from optimal. Our choice of proposal mixtures was based on a few hundred runs using several
different mixtures. There is little doubt that a better algorithm could be constructed that
uses a larger variety of proposal distributions in a more optimal mixture.

The improved F-MCMC algorithm was tested on a variety of simulated data sets that included up to 10 sources
in a 100 bin snippet (once again we are using one year of observations). The algorithm performed very
well, and was able to accurately recover all sources with ${\rm SNR} > 5$ so long as the degree of source
correlation was not too large. Generally the algorithm could de-blend sources that had correlation
coefficients $C_{12} = (h_1\vert h_2)/\sqrt{(h_1\vert h_1)(h_2 \vert h_2)}$ below $0.3$. A full investigation
of the de-blending of highly correlated sources is deferred to a subsequent study. For now we present one
representative example from the 10 source searches. 

\begin{figure}[ht]
\includegraphics[angle=0,width=0.5\textwidth]{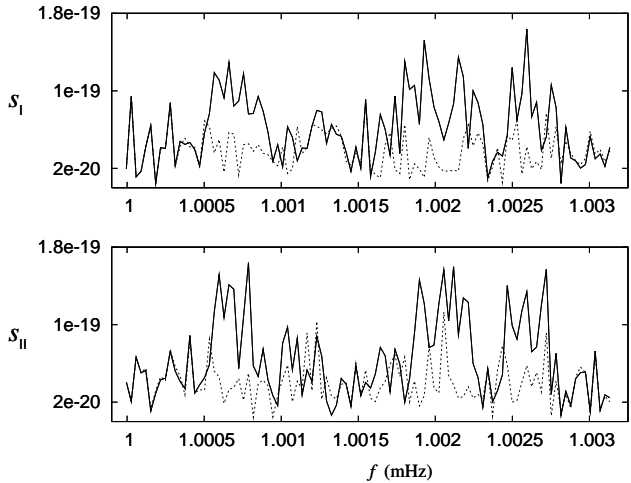}
\caption{\label{pspec}Simulated LISA data with 10 galactic binaries. The solid lines show the
total signal plus noise, while the dashed lines show the instrument noise contribution.}
\end{figure}

\begin{figure}[ht]
\includegraphics[angle=0,width=0.5\textwidth]{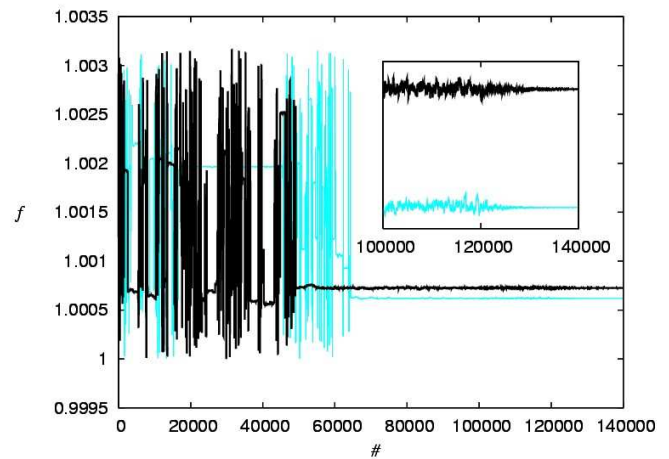}
\caption{\label{burn10}Trace plots of the frequencies for two of the model sources. During the annealing
phase ($\# < 10^5$) the chain explores large regions of parameter space. The inset shows
a zoomed in view of the chain during the MCMC run at $\beta = 1/2$ and the final super cooling which
starts at $\# = 1.2 \times 10^5$.}
\end{figure}

\begin{table}[t]
\caption{F-MCMC search for 10 galactic binaries using a model with 10 sources.
The frequencies are quoted
relative to 1 mHz as $f = 1 \, {\rm mHz} + \delta f$ with $\delta f$ in
$\mu {\rm Hz}$.}
\begin{tabular}{|l|c|c|c|c|c|c|c|c|}
\hline
  & SNR & $A$ ($10^{-22}$) & $\delta f $ & $\theta$ & $\phi$ & $\psi$ & $\iota$ & $\varphi_0$ \\
 \hline 
True & 8.1    & 0.56 & 0.623 & 1.18 & 4.15 & 2.24 & 2.31 & 1.45 \\
MCMC ML & 8.6 & 0.76 & 0.619 & 1.13 & 4.16 & 1.86 & 2.07 & 1.01 \\
 \hline 
True & 9.0     & 0.47 & 0.725 & 0.80 & 0.69 & 0.18 & 0.21 & 2.90 \\
MCMC ML & 11.3 & 0.97 & 0.725 & 0.67 & 0.70 & 0.82 & 0.99 & 1.41 \\
 \hline 
True & 5.1    & 0.46 & 0.907 & 2.35 & 0.86 & 0.01 & 2.09 & 2.15 \\
MCMC ML & 6.0 & 0.67 & 0.910 & 2.07 & 0.61 & 3.13 & 1.88 &  2.28  \\
 \hline 
True & 8.3    & 1.05 & 1.126 & 1.48 & 2.91 & 0.46 & 1.42 & 1.67 \\
MCMC ML & 6.9 & 0.75 & 1.114 & 1.24 & 3.01 & 0.40 & 1.26 & 2.88 \\
 \hline 
True & 8.2    & 0.54 &  1.732 & 1.45 & 0.82 & 1.58 & 0.79 & 2.05 \\
MCMC ML & 7.7 & 0.77 &  1.730 & 1.99 & 0.69 & 1.27 & 1.18 & 2.73 \\
 \hline 
True & 14.7    & 1.16 & 1.969 & 1.92 & 0.01 & 1.04 & 2.17 & 5.70 \\
MCMC ML & 12.8 & 1.20 & 1.964 & 1.97 & 6.16 & 0.97 & 2.00 & 6.15 \\
 \hline 
True & 4.9    & 0.41 & 2.057 &  2.19 & 1.12 & 1.04 & 2.13 & 3.95 \\
MCMC ML & 5.2 & 0.66 & 1.275 &  0.57 & 2.81 & 0.57 & 1.82 & 3.93 \\
 \hline 
True & 8.8     & 0.85  & 2.186 & 2.21 & 4.65 & 3.13 & 2.01 & 4.52 \\
MCMC ML & 10.0 &  1.01 & 2.182 & 2.43 & 5.06 & 0.26 & 2.00 & 5.54 \\
 \hline 
True & 7.6    & 0.58 & 2.530 &  2.57 & 0.01 & 0.06 & 0.86 & 0.50 \\
MCMC ML & 6.7 & 0.98 & 2.582 &  2.55 & 6.03 & 2.71 & 1.52 & 5.58 \\
 \hline 
True & 11.7    & 0.69 & 2.632 & 1.17 & 3.14 & 0.45 & 2.53 & 0.69  \\
MCMC ML & 13.5 & 1.39 & 2.627 & 1.55 & 3.07 & 3.08 & 1.94 & 6.07 \\
\hline 
\end{tabular}
\label{tab10}
\end{table}

A set of 10 galactic sources was randomly selected
from the frequency range $f \in [0.999995,1.003164]$ mHz and their signals were processed through
a model of the LISA instrument response. The root spectral densities in the two independent LISA data
channels are shown in Figure~\ref{pspec}, and the source parameters are listed in Table~\ref{tab10}.
Note that one of the sources had a SNR below 5. The data was then search using our improved F-MCMC
algorithm using a model with 10 sources (70 parameters). The annealing time was set at $10^5$
steps, and this was followed by a short MCMC run of $2\times 10^4$ steps and a super cooling phase
that lasted $2\times 10^4$ steps. The main MCMC run was kept short as we were mostly interested
in extracting maximum likelihood estimates. Figure~\ref{burn10} shows a trace plot of the chain
that focuses on the frequencies of two of the model sources. During the early hot phase the chain
moves all over parameter space, but as the system cools to $\beta=1/2$ the chain settles down and
locks onto the sources. During the final super cooling phase the movement of the chain is
exponentially damped as the model is trapped at a mode of shrinking width and increasing height.

The list of recovered sources can be found in
Table~\ref{tab10}. The low SNR source (${\rm SNR}= 4.9$) was not recovered, but because the model
was asked to find 10 sources it instead dug up a spurious source with ${\rm SNR} = 5.2$. With two
exceptions, the intrinsic parameters for the other 9 sources were recovered to within $3\sigma$
of the true parameters (using the Fisher matrix estimate of the parameter recovery errors).
The two misses were the frequency of the source at $f=1.00253$ mHz (out by $19\sigma$) and the
co-latitude of the the source at $f=1.002632$ mHz (out by $6\sigma$). It is no co-incidence
that these misses occurred for the two most highly correlated sources ($C_{9,10}=-0.23$). The full
source cross-correlation matrix is listed in (\ref{cross}).
\begin{eqnarray}\label{cross}
&& C_{ij} = \frac{(h_i \vert h_j)}{\sqrt{(h_i \vert h_i)(h_j \vert h_j)}} \nonumber \\
&& \nonumber \\
&&=
\left( 
\begin{tabular}{cccccccccc}
{\scriptsize 1 }& {\scriptsize 0.08} & {\scriptsize 0} & {\scriptsize 0.01}
& {\scriptsize 0} & {\scriptsize 0} & {\scriptsize 0} & {\scriptsize 0}
& {\scriptsize 0} & {\scriptsize 0}  \\
{\scriptsize 0.08} & {\scriptsize 1} & {\scriptsize 0.02} & {\scriptsize 0.01} & {\scriptsize 0}
& {\scriptsize 0} & {\scriptsize 0} & {\scriptsize 0} & {\scriptsize 0} & {\scriptsize 0}   \\
{\scriptsize 0} & {\scriptsize 0.02} & {\scriptsize 1} & {\scriptsize -0.06} & {\scriptsize 0} 
& {\scriptsize 0} & {\scriptsize 0} & {\scriptsize 0} & {\scriptsize 0} & {\scriptsize 0}   \\
{\scriptsize 0.01} & {\scriptsize 0.01} & {\scriptsize -0.06} & {\scriptsize 1} & {\scriptsize 0}
& {\scriptsize 0} & {\scriptsize 0} & {\scriptsize 0} & {\scriptsize 0} & {\scriptsize 0}   \\
{\scriptsize 0} & {\scriptsize 0} & {\scriptsize 0} & {\scriptsize 0} & {\scriptsize 1} & {\scriptsize 0}
& {\scriptsize 0} & {\scriptsize 0.01} & {\scriptsize 0} & {\scriptsize 0}   \\
{\scriptsize 0} & {\scriptsize 0} & {\scriptsize 0} & {\scriptsize 0} & {\scriptsize 0} & {\scriptsize 1}
& {\scriptsize -0.03} & {\scriptsize 0.03} & {\scriptsize 0} & {\scriptsize 0}  \\
{\scriptsize 0} & {\scriptsize 0} & {\scriptsize 0} & {\scriptsize 0} & {\scriptsize 0} & {\scriptsize -0.03}
 & {\scriptsize 1} & {\scriptsize -0.05} & {\scriptsize 0} & {\scriptsize 0}  \\
{\scriptsize 0} & {\scriptsize 0} & {\scriptsize 0} & {\scriptsize 0} & {\scriptsize 0.01} & {\scriptsize 0.03}
 & {\scriptsize -0.05} & {\scriptsize 1} & {\scriptsize 0} & {\scriptsize 0}  \\
{\scriptsize 0} & {\scriptsize 0} & {\scriptsize 0} & {\scriptsize 0} & {\scriptsize 0} & {\scriptsize 0}
& {\scriptsize 0} & {\scriptsize 0} & {\scriptsize 1} & {\scriptsize -0.23} \\
{\scriptsize 0} & {\scriptsize 0} & {\scriptsize 0} & {\scriptsize 0} & {\scriptsize 0}
& {\scriptsize 0} & {\scriptsize 0} & {\scriptsize 0} & {\scriptsize-0.23} & {\scriptsize 1}  \\ 
\end{tabular}
\right) \nonumber \\ 
\end{eqnarray}

\begin{figure}[h]
\includegraphics[angle=0,width=0.5\textwidth]{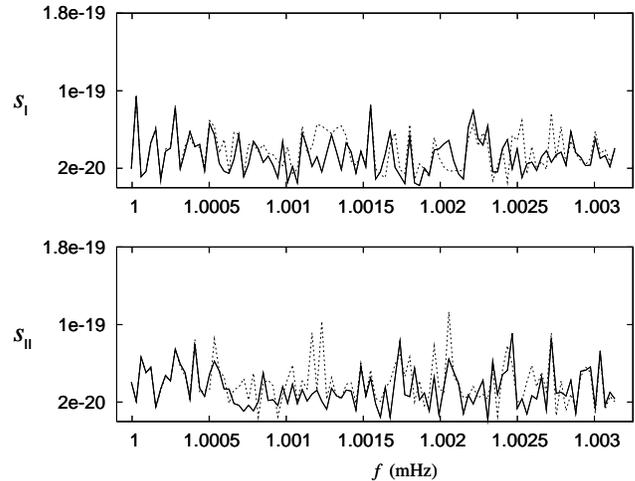}
\caption{\label{pspec_rem}The LISA data channels with the sources regressed using the maximum likelihood
parameter estimates from the F-MCMC search. The solid lines show the residuals,
while the dashed lines show the instrument noise contribution.}
\end{figure}

The MCMC derived maximum likelihood estimates for the the source parameters can be used to
regress the sources from the data streams. Figure~\ref{pspec_rem} compares the residual
signal to the instrument noise. The total residual power is below the instrument noise
level as some of the noise has been incorporated into the recovered signals.

\section{Model Selection}\label{bayes}

In the preceding examples we used models that had the same number of components as
there were sources in the data snippet. This luxury will not be available with the real LISA data.
A realistic data analysis procedure will have to explore model space as well as parameter space. It is
possible to generalize the MCMC approach to simultaneously explore both spaces by incorporating
trans-dimensional moves in the proposal distributions. In other words, proposals that change the number
of sources being used in the fit. One popular method for doing this is Reverse Jump MCMC~\cite{rj}, but
there are other simpler methods that can be used. When trans-dimensional moves are built into the
MCMC algorithm the odds ratio for the competing models is given by the fraction of the time that
the chain spends exploring each model. While trans-dimensional searches provide an elegant solution
to the model determination problem in principle, they can perform very poorly in practice as
the chain is often reluctant to accept a trans-dimensional move.

A simpler alternative is to compare the outputs of MCMC runs using models of fixed dimension.
The odds ratio can then calculated using Bayes factors. Calculating the marginal likelihood of
a model is generally very difficult as it involves an integral over all of parameter space:
\begin{equation}
p_X(s) = \int p(s \vert \vec{\lambda},X) p(\vec{\lambda},X) d\vec{\lambda} \, .
\end{equation}
Unfortunately, this integrand is not weighted by the posterior distribution, so we cannot
use the output of the MCMC algorithm to compute the integral. When the likelihood distribution
has a single dominant mode, the integrand can be approximated using the Laplace approximation:
\begin{eqnarray}
&& p(\vec{\lambda},X) p(s \vert \vec{\lambda}, X) \simeq   p(\vec{\lambda}_{\rm ML},X)
p(s \vert \vec{\lambda}_{\rm ML}, X) \nonumber \\
&& \quad \times  \exp \left( -\frac{ (\vec{\lambda}-\vec{\lambda}_{\rm ML})
\cdot F \cdot (\vec{\lambda}-\vec{\lambda}_{\rm ML})}{2} \right) \, .
\end{eqnarray}
where $F$ is given by the Hessian
\begin{equation}
F_{ij} = \frac{\partial^2 \ln(p(\vec{\lambda},X) p(s \vert \vec{\lambda}, X))}
{\partial \lambda_i \partial \lambda_j}{\Big \vert}_{\vec{\lambda}=\vec{\lambda}_{\rm ML}} \, .
\end{equation}
When the priors $p(\vec{\lambda},X)$ are uniform or at least slowly varying at maximum
likelihood, $F_{ij}$ is equal to the Fisher matrix $\Gamma_{ij}$. The integral is now
straightforward and yields
\begin{equation}\label{lap}
p_X(s) \simeq p(\vec{\lambda}_{\rm ML},X) p(s \vert \vec{\lambda}_{\rm ML}, X) \frac{(2\pi)^{D/2}}{{\rm det}{F}} \, .
\end{equation}
With uniform priors $ p(\vec{\lambda}_{\rm ML},X)$=$1/V$, where
$V$ is the volume of parameter space, and $(2\pi)^{D/2}/{\rm det}{F}$=$\Delta V$, where
$\Delta V$ is the volume of the error ellipsoid.

To illustrate how the Bayes factor can be used in model selection, we repeated the F-MCMC search described
in the previous section, but this time using a model with 9 sources. The results of a typical run are
presented in Table~\ref{tab9}. The parameters of the 9 brightest sources were all recovered to within
$3\sigma$ of the input values, save for the sky location of the source with frequency $f=1.00253$ mHz.
It appears that confusion with the source at $f=1.002632$ mHz may have caused the chain to favour
a secondary mode like the one seen in Figure~\ref{sky}. Using (\ref{lap}) to estimate the marginal
likelihoods for the 9 and 10 parameter models we found $\ln p_9(s) = -384.3$ and $\ln p_{10}(s) = -394.9$,
which gives an odds ratio of $1:4\times 10^4$ in favour of the 9 parameter model. In contrast, a
naive comparison of log likelihoods, $\ln {\cal L}_9 = 413.1$ and $\ln {\cal L}_{10} = 425.7$ would have
favoured the 10 parameter model.

\begin{table}[t]
\caption{F-MCMC search for 10 galactic binaries using a model with 9 sources.
The frequencies are quoted
relative to 1 mHz as $f = 1 \, {\rm mHz} + \delta f$ with $\delta f$ in
$\mu {\rm Hz}$.}
\begin{tabular}{|l|c|c|c|c|c|c|c|c|}
\hline
  & SNR & $A$ ($10^{-22}$) & $\delta f $ & $\theta$ & $\phi$ & $\psi$ & $\iota$ & $\varphi_0$ \\
 \hline 
True & 8.1    & 0.56 & 0.623 & 1.18 & 4.15 & 2.24 & 2.31 & 1.45 \\
MCMC ML & 8.6 & 0.77 & 0.619 & 1.12 & 4.16 & 1.86 & 2.06 & 1.02 \\
 \hline 
True & 9.0     & 0.47 & 0.725 & 0.80 & 0.69 & 0.18 & 0.21 & 2.90 \\
MCMC ML & 11.3 & 0.95 & 0.725 & 0.67 & 0.70 & 0.84 & 0.98 & 1.30 \\
 \hline 
True & 5.1    & 0.46 & 0.907 & 2.35 & 0.86 & 0.01 & 2.09 & 2.15 \\
MCMC ML & 6.2 & 0.75 & 0.910 & 2.09 & 0.61 & 3.09 & 1.82 &  2.21  \\
 \hline 
True & 8.3    & 1.05 & 1.126 & 1.48 & 2.91 & 0.46 & 1.42 & 1.67 \\
MCMC ML & 7.0 & 0.81 & 1.112 & 1.24 & 2.95 & 0.45 & 1.31 & 2.55 \\
 \hline 
True & 8.2    & 0.54 &  1.732 & 1.45 & 0.82 & 1.58 & 0.79 & 2.05 \\
MCMC ML & 7.5 & 0.76 &  1.730 & 1.95 & 0.70 & 1.23 & 1.19 & 2.68 \\
 \hline 
True & 14.7    & 1.16 & 1.969 & 1.92 & 0.01 & 1.04 & 2.17 & 5.70 \\
MCMC ML & 12.9 & 1.23 & 1.965 & 1.97 & 6.17 & 0.99 & 1.99 & 6.11 \\
 \hline 
True & 4.9    & 0.41 & 2.057 &  2.19 & 1.12 & 1.04 & 2.13 & 3.95 \\
MCMC ML & - & - & - &  - & - & - & - & - \\
 \hline 
True & 8.8     & 0.85  & 2.186 & 2.21 & 4.65 & 3.13 & 2.01 & 4.52 \\
MCMC ML & 10.0 &  1.00 & 2.182 & 2.41 & 5.05 & 0.23 & 2.00 & 5.59 \\
 \hline 
True & 7.6    & 0.58 & 2.530 &  2.57 & 0.01 & 0.06 & 0.86 & 0.50 \\
MCMC ML & 5.8 & 0.72 & 2.536 &  0.58 & 5.70 & 3.04 & 1.52 & 4.64 \\
 \hline 
True & 11.7    & 0.69 & 2.632 & 1.17 & 3.14 & 0.45 & 2.53 & 0.69  \\
MCMC ML & 13.2 & 1.21 & 2.631 & 1.41 & 2.97 & 0.46 & 2.02 & 0.50 \\
\hline 
\end{tabular}
\label{tab9}
\end{table}

It is also interesting to compare the output of the 10 source MCMC search to
the maximum likelihood one gets by starting at the true source parameters
then applying the super cooling procedure (in other words, cheat by starting
in the neighborhood of the true solution). We found $p_{\rm cheat}(s)=-394.5$,
and $\ln {\cal L}_{\rm cheat} = 421.5$, which tells us that the MCMC solution,
while getting two of the source parameters wrong, provides an equally good
fit to the data. In other words, there is {\em no} data analysis algorithm
that can fully deblend the two highly overlapping sources.

\section{Conclusion}\label{conclude}

Our first pass at applying the MCMC method to LISA data analysis has shown the
method to have considerable promise. The next step is to push the existing
algorithm until it breaks. Simulations of the galactic background suggest that
bright galactic sources reach a peak density of one source per five $1/{\rm year}$
frequency bins~\cite{seth}. We have shown that our current F-MCMC algorithm can handle
a source density of one source per ten frequency bins across a one hundred bin snippet.
We have yet to try larger numbers of sources as the current version of the algorithm employs
the full $D=7N$ dimensional Fisher matrix in many of the updates, which leads to a large
computational overhead. We are in the process of modifying the algorithm so that sources are
first grouped into blocks that have strong overlap. Each block is effectively independent
of the others. This allows each block to be updated separately, while still taking care of
any strongly correlated parameters that might impede mixing of the chain. We have already
seen some evidence that high local source densities pose a challenge to the current
algorithm. The lesson so far has been that adding new, specially tailored proposal distributions
to the mix helps to keep the chain from sticking at secondary modes of the posterior (it takes a
cocktail to solve the cocktail party problem). On
the other hand, we have also seen evidence of strong multi-modality whereby the secondary
modes have likelihoods within a few percent of the global maximum. In those cases the chain
tends to jump back and forth between modes before being forced into a decision by the
super-cooling process that follows the main MCMC run. Indeed, we may already be pushing
the limits of what is possible using any data analysis method. For example, the 10 source
search used a model with 70 parameters to fit 400 pieces of data (2 channels $\times$ 2 Fourier
components $\times$ 100 bins). One of our goals is to better understand the theoretical limits
of what can be achieved so that we know when to stop trying to improve the algorithm!

It would be interesting to compare the performance of the different methods that have
been proposed to solve the LISA cocktail party problem. Do iterative methods like
gCLEAN and Slice \& Dice or global maximization methods like Maximum Entropy
have different strengths and weakness compared to MCMC methods, or do they all fail
in the same way as they approach the confusion limit? It may well be that methods that
perform better with idealized, stationary, Gaussian instrument noise will not prove
to be the best when faced with real instrumental noise.

\begin{acknowledgements}
This work was supported by NASA Cooperative Agreement NCC5-579.
\end{acknowledgements}

\end{document}